Response of Atmospheric Biomarkers to $NO_x$-induced Photochemistry Generated by Stellar Cosmic Rays for Earth-like Planets in the Habitable Zone of M-Dwarf Stars

*John Lee Grenfell[1\*], Jean-Mathias Grießmeier[2#], Philip von Paris[3$], A. Beate C. Patzer[1], Helmut Lammer[4], Barbara Stracke[3], Stefanie Gebauer[1], Franz Schreier[5], and Heike Rauer[1,3]*

(1) Zentrum für Astronomie und Astrophysik
Technische Universität Berlin (TUB)
Hardenbergstr. 36
10623 Berlin
Germany

(2) ASTRON
PO Box 2
7990 AA Dwingeloo
The Netherlands

(3) Institut für Planetenforschung
Deutsches Zentrum für Luft- und Raumfahrt (DLR)
Rutherford Str. 2
12489 Berlin
Germany

(4) Österreichische Akademie der Wissenschaften
Schmiedlstr. 6
8042 Graz
Austria

(5) Institut für Methodik der Fernerkundung
Deutsches Zentrum für Luft- und Raumfahrt (DLR)
Oberpfaffenhofen
82234 Wessling
Germany

# Now at:
Laboratoire de Physique et Chemie de l'Environment et de l'Espace (LPC2E)
Université d'Orléans / CNRS
3A, Avenue de la Recherche Scientifique
45071 Orleans cedex 2
France

$ Now at:
(i) Univ. Bordeaux, LAB, UMR 5804, F-33270, Floirac, France

(ii) CNRS, LAB, UMR 5804, F-33270, Floirac, France

\* Corresponding author

*Abstract: Understanding whether M-dwarf stars may host habitable planets with Earth-like atmospheres and biospheres is a major goal in exoplanet research. If such planets exist, the question remains as to whether they could be identified via spectral signatures of biomarkers. Such planets may be exposed to extreme intensities of cosmic rays that could perturb their atmospheric photochemistry. Here, we consider stellar activity of M-dwarfs ranging from quiet up to strong flaring conditions and investigate one particular effect upon biomarkers, namely, the ability of secondary electrons caused by stellar cosmic rays to break up atmospheric molecular nitrogen ($N_2$), which leads to production of nitrogen oxides ($NO_x$) in the planetary atmosphere, hence affecting biomarkers such as ozone ($O_3$). We apply a stationary model, that is, without a time-dependence, hence we are calculating the limiting case where the atmospheric chemistry response time of the biomarkers is assumed to be slow and remains constant compared with rapid forcing by the impinging stellar flares. This point should be further explored in future work with time-dependent models. We estimate the $NO_x$ production using an air shower approach, and evaluate the implications using a climate-chemical model of the planetary atmosphere. $O_3$ formation proceeds via the reaction $O+O_2+M \rightarrow O_3+M$. At high $NO_x$ abundances, the O atoms arise mainly from $NO_2$ photolysis, whereas on Earth this occurs via the photolysis of molecular oxygen ($O_2$). For the flaring case $O_3$ is mainly destroyed via direct titration: $NO+O_3 \rightarrow NO_2+O_2$, and not via the familiar catalytic cycle photochemistry, which occurs on Earth. For scenarios with low $O_3$, Rayleigh scattering by the main atmospheric gases ($O_2$, $N_2$, and $CO_2$) became more important for shielding the planetary surface from ultra-violet radiation. A major result of this work is that the biomarker $O_3$ survived all the stellar-activity scenarios considered except for the strong case, whereas the biomarker nitrous oxide ($N_2O$) could survive in the planetary atmosphere under all*

*conditions of stellar activity considered here, which clearly has important implications for missions that aim to detect spectroscopic biomarkers.*

**1. Introduction**

M-dwarf stars are numerous, long-lived, and have relatively high planet-to-star luminosity ratios. Therefore, they have been suggested as promising targets for searches of habitable planets (e.g., Scalo et al., 2007; Rauer et al., 2011). Such systems will, however, possess a close-in Habitable Zone (HZ) where, for example, coronal mass ejections (CMEs), which are frequent for active M-dwarf stars, may erode the planetary atmosphere (Khodachenko et al., 2007; Lammer et al. 2007, 2008). For a young planet, the dense and fast stellar wind can play a similar role (Grieβmeier et al., 2010). Clearly, for life to develop habitable conditions must be maintained for sufficiently long timescales. More modeling studies of, for example, atmospheric escape and the long chemical development of the atmosphere are required to gain more understanding. Here, we concentrate on the effect of high Stellar Cosmic Rays (SCRs) on biomarker molecules of terrestrial exoplanets.

The particle flux of Galactic Cosmic Rays (GCRs) to Earth-like planets in the HZ around M-dwarf stars was studied by Grießmeier et al. (2005, 2009), and resulting effects on atmospheric chemistry were discussed by Grenfell et al. (2007)[1] and Grießmeier et al. (2009, 2010). In the present work, we extend these studies and look at the effect of SCRs caused by stellar activity ranging from quiet up to strong flaring conditions on the atmospheric chemistry. A motivation for this work is that SCRs could be especially dominant compared with GCRs for young, active M-dwarf stars. In these cases, not only are SCRs especially significant, but furthermore the strong stellar magnetic fields lead to strong attenuation of the GCRs. We focus on the particle effects under differing conditions. In a recent work, Segura et al. (2010) investigated the photochemistry induced on such Earth-like atmospheres by modeling the high energy radiation and charged particle effects associated with a single, strong flare. They showed that stellar UV radiation emitted during the flare does not significantly affect the ozone column depth of the planet. We build on their work by assuming a harder criterion, namely, constant flaring conditions (see below) for our considered planetary atmosphere. Also, whereas the Segura work estimated $NO_x$-production from observed X-ray proxies, in our work we applied a theoretical air-shower approach to determine such $NO_x$-production directly (as we will show both approaches compare quite well). Finally, compared with the Segura et al. work we build on and extend the chemical analysis, providing new insight into the responses of biomarkers, related species, and UV protection of the planetary surface. Section 2 describes the models, section 3 presents results, and section 4 draws conclusions with regard to the response of atmospheric biomarkers and their spectral signatures.

## 1.1 Cosmic Rays (CRs) and Magnetic Fields

GCRs travel from their source regions (e.g., supernovae) through the interstellar medium (ISM) and penetrate the stellar magnetic field (for a good review see Scherer et al. 2002). Heavier CR particles consisting of, for example, helium, carbon, and iron nuclei (Gloeckler et al. 2009; Ziegler, 1996; Seo et al. 1991) are known, but their fluxes are lower by at least an order of magnitude compared with the proton fluxes. High energy SCRs can be richer in heavier elements (Smart and Shea, 2002). Our work however, assumes the CR particles are protons.

Penetration of CRs through a planet's protective magnetosphere and associated effects are the subject of a rich body of literature. Melott and Thomas (2011) and Dartnell (2011) and references therein provided reviews. Both heliosphere and magnetosphere fields are variable in time and space. In an exoplanetary context, a planet with an Earth-like magnetic field will likely direct CRs with modest energies (up to several hundreds of MeV) to its polar regions, as is the case for the Auroral zones of Earth (see Tarduno et al., 2010 for a review), whereas for planets with weak or no magnetic fields, high energy CRs could also impinge at lower latitudes.

**1.2 Interactions of CRs with Atmospheres and Biomarkers**

Cosmic Rays interact with planetary atmospheres in a range of ways. They can induce ionization as observed, for example, on Titan and Venus (Dartnell, 2011) and in association with lightning (e.g., Dwyer, 2005), acid rain, and cloud formation on Earth (e.g., Svensmark and Christensen, 1997). A 3D model would be required for a full description, although a 1D study is an adequate first step, since the necessary boundary conditions for a 3D study are completely lacking for extrasolar planets. On Earth, $O_3$ features a seasonal cycle where winter/spring column values (Brasseur, et al., 1999) are enhanced by about 50% due to weak photolytic loss in winter. It has a latitudinal gradient being produced in the Earth's Tropics and transported by the meridional circulation to the pole where its column value peaks. At cold polar latitudes in winter, $O_3$ is rapidly lost via fast catalytic chlorine cycles (e.g., WMO, 1994). Ozone varies in altitude - in Earth's troposphere it is produced by the smog mechanism (Haagen-Smit, 1952). In Earth's stratosphere, $O_3$ is produced by the Chapman mechanism (Chapman, 1930). In the mesosphere and thermosphere, where $O_3$ is affected by fast hydrogen-oxide loss cycles and direct photolytic loss, it displays a diurnal cycle. A discussion and implications for future 3D model studies is presented in the discussion.

Solar Proton Events (SPEs) can form $NO_x$ (from $N_2$) and $HO_x$ (from $O_2$), which affect $O_3$ via chemical reactions. $NO_x$ (in the form of NO) formation on Earth from SPEs occurs typically at (70-100) km – here it can be quickly photolysed, but in winter it survives in the dark polar vortex to be transported slowly (over periods of typically several weeks) down to the mid-stratosphere where the $O_3$ layer resides. In spring, the vortex breaks up and the enhanced $NO_x$ is transported to lower latitudes where it encounters sunlight, which initiates catalytic $O_3$ loss cycles. The removal timescales of $NO_x$ and $HO_x$ are typically a couple of weeks in Earth's mid-stratosphere. In Earth's troposphere, at very high $NO_x$ abundances of about a few ppm, $O_3$ is removed directly. At more modest $NO_x$ abundances of about 1000 times lower, smog $O_3$ is produced.

SPE events on Earth have been documented to perturb stratospheric $NO_x$ at high-latitudes and, hence, may have a significant impact upon $O_3$. For example, the large 1989 SPE led to a (1-2%) decrease in the $O_3$ column averaged over $50^o$ to $90^o$N (Jackman et al. 2000), whereas the significant 2003 SPE led to a lowering in local $O_3$ by several tens of percent in the mid to upper stratosphere (Jackman et al. 2005). In recent years, numerous 3D photochemical modeling studies of SPEs have been applied to Earth's atmosphere, for example, Jackman et al. (2011) and references therein simulated charged particle precipitation during the January 2005 ground level event, calculating $NO_x$ abundances of greater than 50 parts per billion by volume (ppbv) in the polar mesosphere. Satellite observations were also simultaneously employed to study the perturbed atmospheric composition and the relaxation of active oxides into key reservoir molecules such as $HNO_3$ and $H_2O_2$.

## 2. Description of Model Approach, Model and Runs

**Model Approach** - First, we calculate the Top-of-Atmosphere (TOA) global mean SCR fluxes based on observations of solar particles taken above Earth's magnetosphere for an unmagnetized planet located in the HZ around an M-dwarf star (MHZ). Second, these SCR fluxes induce an enhanced $NO_x$ production in the planetary atmosphere as described by the air shower approach in the radiative convective photochemical column model (section 2.1). Finally, the resulting chemical concentration and temperature profiles from the column model are input into a line-by-line spectral model to calculate theoretical emission spectra (section 2.2).

### 2.1 Coupled Climate-Chemistry Atmospheric Column Model

We use the new model version of the coupled chemistry-climate 1D atmospheric model of Rauer et al. (2011) and references therein. Recent model updates since the work of Grenfell et al. (2007)[1,2] include a new offline binning routine for calculating the input stellar spectra and a variable vertical atmospheric height in the model (more details are given in Rauer et al. (2011). There are two main modules, namely, a radiative-convective climate module and a chemistry module.

*The Climate Module* is a global-mean stationary, hydrostatic column model of the atmosphere. The model pressure grid ranges from the surface up to $6.6 \times 10^{-5}$ bar (for the Earth this corresponds to a height of about 70km). Initial composition, pressure, and temperature profiles are based on modern Earth. The radiative transfer calculation is based on the RRTM (Rapid Radiative Transfer Module) for the thermal radiation. This scheme applies 16 spectral bands by using the correlated k-method for the major atmospheric absorbers. The RRTM validity range (see Mlawer et al. 1997) for a particular altitude corresponds to Earth's modern climatological mean temperature ±30K for pressures from ($10^{-5}$ to 1.05) bar and for a $CO_2$ abundance from modern-day up to x100 that value. The shortwave radiation scheme consists of 38 spectral intervals for the major absorbers and includes Rayleigh scattering for $N_2$, $O_2$, and $CO_2$ with cross-sections based on Vardavas and Carver (1984). A constant, geometrical-mean, solar-zenith angle = $60^o$ is used. In the troposphere, a moist adiabatic lapse rate is assumed where the radiative temperature gradient is larger than the moist adiabatic gradient. Humidity in the troposphere is based on Earth observations (Manabe and Wetherald, 1967). For the Earth-around-the-Sun scenario, a solar spectrum based on Gueymard (2004) is employed. For the M-dwarf scenarios, an AD Leo stellar spectrum is derived from observations by Pettersen and Hawley (1989) as taken from the IUE satellite and observation by Leggett et al. (1996) in the near-IR and based on a nextGen stellar model spectrum for wavelengths beyond 2.4 microns (Hauschildt et al., 1999). The climate effects of clouds are not included explicitly, although these are considered in a straightforward way by adjusting the surface albedo to a value of 0.21 in all runs, that is, for which the mean surface temperature of Earth (288 K) was attained for the Earth control run. Starting values (T,p) for the climate module were based on the US-standard atmosphere (1976). The climate module ran to convergence, then outputs temperature, pressure, and water abundances were taken as input for the chemistry module.

*The Chemistry Module* was described by Pavlov and Kasting et al. (2002). The current scheme includes 55 chemical species for 212 reactions with chemical kinetic data taken from the Jet Propulsion Laboratory (JPL) (2003) Report. The kinetic data are typically measured from about (250-400)K, although each reaction has its own stated range.

Three-body reactions are usually measured in an $N_2$-$O_2$ bathgas (e.g., JPL Report 14, 2003). Note that different rate constants by up to about a factor 10 may be expected for reactions in, for example, $CO_2$-dominated atmospheres, although we do not consider these here. The chemistry module assumes a planet with an Earth-like development, that is, with an $N_2$-$O_2$ dominated atmosphere, etc. The reaction scheme is designed to reproduce modern Earth's atmospheric composition with a focus on biomarkers (e.g., $O_3$, $N_2O$) and major greenhouse gases such as $CH_4$. It therefore includes fast nitrogen-oxide and hydrogen-oxide reactions and their reservoir species, reproducing the Earth's atmospheric $O_3$ abundance, as well as the major sources and sinks of hydroxyl (OH), reproducing the atmospheric $CH_4$ profile. The chemical module calculates the steady-state solution of the usual 1D continuity equations by an implicit Euler method. Atmospheric mixing between adjacent gridboxes is parameterized via Eddy diffusion coefficients (K). K is set to $10^5$ cm$^2$ s$^{-1}$ from the surface up to z=10km, decreasing to K=$10^{3.5}$ cm$^2$ s$^{-1}$ at z~16km then increasing up to K=$10^{5.5}$ cm$^2$ s$^{-1}$ at z~64km. From a total of 55 chemical species, 34 are "long-lived," that is, their concentrations are obtained by solving the full continuity equation. A further 16 species are "short-lived," that is, their concentrations are calculated exclusively from the long-lived species concentrations by assuming chemical steady-state, that is, equating production rates with loss rates for each short-lived species. Finally, the remaining, so-called "isoprofile species" are set to constant abundances in all gridboxes, namely, $CO_2$ = 3.55x10$^{-4}$, $O_2$ = 0.21 and $N_2$ ~0.78 volume mixing ratio (vmr).

Boundary conditions in the chemistry module were based on reproduction of the modern Earth atmospheric composition and included constant surface biogenic ($CH_3Cl$, $N_2O$) and source gas ($CH_4$, CO) species (see Grenfell et al., 2011 for more details). Also included are modern-day tropospheric lightning emissions of nitrogen monoxide (NO), volcanic sulphur emissions of $SO_2$ and $H_2S$, and a constant effusion flux of CO and O at the upper model boundary, which represents the photolysis products of $CO_2$. Surface removal of long-lived species via dry and wet deposition is included via deposition velocities (for dry deposition) and Henry's law constants (for wet deposition). The chemistry module ran to convergence and then output its greenhouse gas abundances to the climate module. This whole process, that is, exchange of data between the two modules, is repeated until the chemical concentrations and the atmospheric temperature and pressure structure converge.

*The Cosmic Ray Scheme* - Starting with the incoming TOA time-average proton fluxes calculated by the Cosmic Ray Proton Flux (CRPF) model (Grieβmeier et al., 2005) (Figure 1), we calculated NO (hence $NO_x$) production in our atmospheric column model as in Grenfell et al. (2007)[1]. We consider thereby only incoming proton energies of 64MeV and above. Somewhat lower energies down to ~10 MeV may also affect Earth's atmosphere but mostly at altitudes above the mesosphere, not considered in our work. The CRPF model was originally designed to simulate the detailed passage of a large number of high energy proton trajectories through Earth's magnetosphere. In its current implementation, the scheme is not suitable for calculating the above-mentioned lower energies. Adding these would increase $NO_x$ generation and $O_3$ loss, so our results can be regarded as conservative in this sense.

The impinging protons produce secondary electrons ($e^-$), hence, $NO_x$ via:

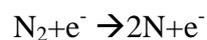

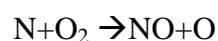

The resulting NO fluxes were then calculated interactively in the atmospheric column model. An important parameter (k) is defined as the number of atomic nitrogen atoms produced on average when one nitrogen molecule is destroyed. Values of k have been suggested in the range k=0.96 (e.g., Nicolet, 1975) up to k=1.25 (e.g., Porter et al., 1976; Jackman et al., 1980). We assumed one $NO_x$ produced per $N_2$ destroyed at all heights (i.e., k=1). We further assumed a third-power law for the secondary electron energy spectrum (Bichsel et al., 2005) with a proton penetration profile based on Gaisser and Hillas (1977) (their Figure 2). For proton energies at and above 256 MeV, we assumed a proton attenuation length (PAL) of 80 g/cm$^2$, and we further assumed a $PAL_{max}$ = 150 g cm$^{-2}$ (defined as the column mass obtained by integrating from the TOA down to the altitude where the shower has its maximum). For the three lower proton energy levels of 256, 128, and 64 MeV, which are important for stellar particles (see Figure 1), we assumed PAL = $10^{-3}$, 1, 2 g/cm$^2$, $PAL_{max}$ = $10^{-2}$, 5, 25 g/cm$^2$, respectively (Amsler et al. 2008). We assumed an electron attenuation length of 40 g/cm$^2$ for all energy levels, also based on Amsler et al. (2008). The bulk of measured PAL data for Earth's atmosphere is obtained at energies at least several orders of magnitude higher. In our study, values were therefore obtained by extrapolation. The Particle Data Group (PDG) (2002) report suggested that the PAL and $PAL_{max}$ values quoted for the three lowest energy levels in our work are uncertain by around a factor 10. Therefore, $NO_x$-catalysed $O_3$ loss calculated in our work may be an underestimate. This caveat therefore is not likely to change the conclusions for the flaring run, where all $O_3$ is destroyed. We performed a parameter variation where PAL values were decreased by up to two orders of magnitude. As a result, the maximum of our CR-induced $NO_x$ production moving from about 15km height up to about 22km (not shown).

It is not possible to validate a stationary model in terms of, for example, $O_3$ depletion by using SPE measurements for Earth since these results are time-dependent. However, we present here a comparison of ion-pair production. Jackman et al. (2005) (their Figure 1) reported peak ion pair production for Earth of about $10^4$ ions cm$^{-3}$ s$^{-1}$ at (60-70)km dropping to about $10^2$ ions cm$^{-3}$ s$^{-1}$ at ~30km based on GOES-11 observations of the October 2003 SPE. By comparison, our column model (SPE Earth conditions) calculated $(2-4) \times 10^2$ ions cm$^{-3}$ s$^{-1}$ above 60km, ~$5 \times 10^4$ at 30km and a peak production of ~$7 \times 10^4$ ions cm$^{-3}$ s$^{-1}$ in the lower stratosphere. In summary, our stationary model over-estimates peak ion production for Earth's mid atmosphere since we simulate constant flaring conditions, although this shortcoming is expected to be less important for the M-star constant flaring case. In a different approach, Segura et al. (2010) aimed to model a transient single flare from AD Leo – the resulting $NO_x$ production rates are, however, rather similar to our values (as discussed later in text).

## 2.2 Theoretical line-by-line (LBL) Spectral Model

The SQuIRRL (Schwarzschild Quadrature InfraRed Radiation Line-by-line) code (Schreier and Schimpf, 2001) was originally designed for high resolution radiative transfer modeling in the IR assuming a spherical atmosphere (for an arbitrary observation geometry, instrumental field-of-view and spectral response function). Assuming local thermodynamic equilibrium (LTE), a Planck function (with appropriately level temperatures) is used as the emission source. The simulated atmosphere is assumed to be cloud and haze free without scattering. SQuIRRL has been verified via extensive intercomparisons with other radiative transfer codes (e.g., von Clarmann et al., 2002; Melsheimer et al., 2005). By using molecular spectroscopic line parameters from the HITRAN 2004 (Rothman et al., 2005) database, absorption coefficients are computed with continuum corrections. The emission spectra are calculated assuming a pencil beam at a viewing angle of 38 degrees as adopted, for example, by Segura et al. (2003).

## 2.3 Application of a stationary SCR model to active M-dwarf star conditions

A stationary atmospheric column model approach assumes that the biomarker chemical responses are slow compared with the stellar variability. Therefore, we now discuss and compare biomarker photochemical timescales with stellar variability timescales for AD-Leo. Note, in our model study we consider only NOx production from CRs (and not the effect of the change in UV-during flaring conditions) and its influence upon the biomarkers. Nevertheless, we briefly discuss stellar (UV) variability of M-dwarf stars in order to investigate the applicability of our time-independent model.

We perform a timescale analysis where we make the assumption that the relevant removal timescale for perturbed NOx (induced by cosmic rays) occurs via oxidation into nitric acid via the reaction $NO_2+OH+M \rightarrow HNO_3+M$, where M, the total atmospheric number density, denotes a 3rd body as a collision partner to remove excess vibrational energy. We also assume that the initial NOx forcing (enhancement via CRs) occurs only once and is immediately followed by removal via the oxidation reaction - this is also a reasonable assumption for a strong SPE in Earth's stratosphere. We note, however, that for M-dwarf scenarios the above assumptions should be further explored by time-dependent models. For example, the UV environment of the planet's atmosphere will influence how NOx is removed, for example, for changed UVB environments; other NOx reservoirs (and not $HNO_3$ as on Earth) could play a greater role in NOx removal. Regarding NOx formation via CRs, the forcing timescale will depend also on, for example, the magnitude and frequency of the stellar flares. For example, for an Earth-like planet orbiting in the HZ around ADLeo, typical flares there would be considered large (in terms of total flare energy) compared to those observed on Earth and may occur 3-4 times per Earth day (see following discussion). However, there also exist higher energy flare events that occur every ~(1-2) Earth weeks (e.g., Güdel et al., 2003, their Figure 3). Although with the above caveats in mind, the following timescale analysis, however, assumes as a first estimate an Earth-like NOx relaxation into $HNO_3$ with a single NOx-forcing event, as described in the following section.

**Biomarker photochemical timescales** - We focus in the mid-stratosphere at z=30km since this is where the $O_3$ layer peaks. Defining the NOx removal timescale, $\tau_{NO2}$ (30km) = $[NO_2]/k[OH][NO_2]$ gives column model values of ~12 days (~43 days) for our Earth-orbiting-the-Sun scenario (Earth-like planet orbiting AD-Leo, flaring scenario). The factor 3.6 increase here reflects the lowering in [OH] for M-dwarf star scenarios, due to the weaker UVB output of the central star, as already discussed, for example, by Segura et al. (2005). The above timescales are probably, however, underestimates, since the resulting $HNO_3$ can photolyze to release its NOx back into the atmosphere. Then the effective timescale for NOx removal can be much longer - up to a few years. For example, Thomas et al. (2005a, 2005b) and Ejzak et al. (2007) showed that atmospheric $O_3$ depletion from enhanced NOx (from simulated, close-by gamma ray bursts) lasts for a long time. $O_3$ depletion is constant for ~2 years, and even after 5-7 years, 10% depletion remains. The Ejzak et al. study further suggests that strong ionizing events led to a ~(33-37)% maximum decrease in the $O_3$ column. In summary, we estimate biomarker timescales of (Earth) months to years in the mid stratosphere, decreasing in the upper stratosphere to mesosphere down to several Earth weeks.

**AD-Leo variability timescales** - Khodachenko et al. (2007) suggested that AD-Leo emits more than 36 CMEs per day. Regarding the stellar photon flux variability, the total energy of the flare in our flaring scenario corresponds to $1.2 \times 10^{32}$ ergs (see e.g., Gopalswamy, 2006), assuming quadratic scaling of CRs with the planet at 0.153AU. For such flares, this energy corresponds to a flare frequency for AD Leo (see Audard et al. 2000) of several per day, consistent with Houdebine et al. (1990). However, the Audard study (their Figure 1) also indicated isolated, very large flaring events occurring on timescales of up to 1-2 Earth weeks, as already mentioned (for more details see also Güdel et al. 2003). For such large flaring events, the stationarity assumption of our model, especially in the upper layers, could be weak. A preliminary analysis of our model results for the AD Leo "flaring run," however, suggested that removal of NOx into $HNO_3$ is still an important chemical process in the stratosphere and that the removal is slower than on Earth. However, it is interesting to revisit this issue with a time-dependent model, and to calculate the resulting timescales for the removal of NOx into its photochemical reservoir molecules.

**Diurnal $O_3$ Observations on Earth** - It is well known that $O_3$ diurnal variations are low in Earth's middle atmosphere. This suggests that the $O_3$ response to changes in UV on timescales of ~12 hours are negligible, which supports the case for applying a stationary model at least for these particular conditions. Clearly, this depends on the planetary rotation rate and is no longer valid for faster rotating planets. We expect the rotation rate of planets considered in this work to be comparable to, or lower than, that of Earth, as discussed.

### 2.4 About the Runs

In this study, we assumed that our hypothetical Earth-like planet is positioned in the MHZ at 0.153AU, the distance where integrated radiative energy fluxes of our adopted input stellar (AD Leo) spectrum are equal to the solar constant.

**Run 1:** Earth orbiting the Sun at 1 AU distance (solar spectrum) conditions, that is, with biomass emissions reproducing modern Earth's biosphere and with TOA galactic cosmic ray (GCR) fluxes as calculated by the CRPF model (see section 2.1 and Grießmeier et al., 2005 for more details) for the magnetized planet Earth (***Earth GCRs run***).

**Run 2**: Same as run 1 but with M-dwarf stellar spectrum (AD Leo) (***GCR M-dwarf star run***)

**Run 3:** Same as run 2 but with GCR input replaced with interplanetary SCRs for an unmagnetised planet based on Kuznetsov et al. (2005) scaled to 0.153 AU by using particle observations over approximately one year of the Sun during the *minimum* of the 11-year solar cycle (***Quiescent M-dwarf star SCR run***).

**Run 4:** Same as run 3 but using particle observations of the Sun during the *maximum* of the 11-year solar cycle (***Active M-dwarf star SCR run***)

**Run 5:** Same as run 4 but with SCRs based on Smart and Shea (2002) scaled to 0.153 AU by using peak particle observations of the Sun during a Solar Proton Event (***Flaring M-dwarf star SCR run***).

**Run 6**: as for run 5 but with a reduced Eddy mixing coefficient by a factor 10.

**Run 7**: as for run 5 but neglecting Rayleigh scattering.

**CR Input Data for the Runs** - For the quiet and active star scenarios, we employed the spectra of Kuznetsov et al. (2005) at 1AU (their Figure 2). For the flaring M-dwarf star case, we used data measured at 1AU by the spacecrafts GOES 6 and 7 on 30th September 1989 (Smart and Shea, 2002, their Figure 11). Particle fluxes are determined by a small number of events with very high values, which makes averaging challenging. The number of events varies with the solar cycle, with higher fluences for about 7 years, followed by a quieter period of about 4 years (Feynman et al., 1993, their Figures 1, 2). However, the fluence during the latter period is not negligible, so that the division into two such categories is debated (Kuznetsov et al. 2001, 2002). These spectra do not correspond to a continuous particle stream from the Sun, but to average over all small, medium, and large events that may happen during one year.

The variation of the stellar cosmic ray flux as a function of distance from the source is not known for stars other than the Sun. For this reason, we base the distance-dependence of the particle flux on that observed for the Sun. Solar particle fluxes have a strong maximum near the region where they are generated. As they travel outwards, this feature is diluted. This results in a longer event duration at large orbital distances and reduces the instantaneous flux during the event due to particle conservation. The effect is stronger for lower energy particles, as seen in observations (Hamilton, 1977, their Figure 10) and in numerical simulations (Hamilton et al., 1990, their Figure 1; Lario et al., 2000, their Figure 9). Feynman and Gabriel (1988) established a first set of power-laws at distance, d < 1 AU for the *peak flux*, $f_{peak}$ and for the *event fluence, $fl_{event}$* (i.e. number of particles integrated over the whole time of the event):

$$f_{peak}(d) = f_{peak}(1AU)*(d/1AU)^\alpha$$

$$fl_{event}(d) = fl_{event}(1AU)*(d/1AU)^\beta$$

where $\alpha$ = (-3 to -2), $\beta$ = (-3 to -2). In the range d = (0.65 to 4.92) AU, Hamilton et al. (1990) suggested that for particles with energies of (10 to 20 MeV), $\alpha$ = (-2.9 to -3.7) and $\beta$ = (-1.8 to -2.4), which is in good agreement with their theoretical model. More recently, Lario et al. (2006) suggested for d < 1 AU (based on 72 observations by Helios 1, Helios 2 and the Eighth Interplanetary Monitoring Platform (IMP-8)), that $\alpha$ = (-2.7 to –1.9) and $\beta$ = (-2.1 to -1.0). They also stated that previous models did not take into account a focusing effect, which is important for d < 1AU. From numerical modeling, Ruzmaikin et al. (2005) concluded that $\alpha$ is energy-dependent, that is, $\alpha$ = -2.6 at 10 MeV, which becomes more negative at higher energies, that is, $\alpha$ = -2.9 at 100 MeV (see their Figure 4). Lario et al. (2006) on the other hand suggested a more negative $\alpha$ at lower energies. Considering the uncertainties, we will use an energy-independent value. Since in our steady-state study, the more relevant quantity is the event fluence and not the peak flux because our model converges to a steady-state. We therefore used the fluence equation above and adopted a mean value for $\beta$ = -2.0.

## 3. Results

### 3.1 Top-of-atmosphere CR Energy Spectra and Nitrogen Monoxide (NO) Production

Figure 1 shows the energy spectrum for primary particles impinging at the top of the planetary atmosphere calculated by the CRPF model. The Earth-orbiting-the-Sun-with-GCRs (run 1, hereafter "Earth GCR") simulation shows enhanced fluxes at the higher energies, whereas the SCR fluxes in the other scenarios dominate the lower energy-regime. Figure 2 shows the number of secondary electrons relative to the shower maximum as a function of incoming proton energy for run 1 (Earth GCR). Above 512 MeV, the energy curves overlap because input parameters (e.g., the maximum attenuation length, $X_{max}$) do not vary over this energy range. The high energy (>512 MeV) cosmic rays penetrate deepest into the atmosphere, leading to secondary electron shower production with values peaking at ~0.5 in the upper troposphere at about 0.15 bar. With decreasing energy of the incoming protons, atmospheric penetration becomes weaker, and the peak value shifts to higher altitudes, that is, in the stratosphere and mesosphere.

Figure 3 shows the nitrogen monoxide (NO) atmospheric production rates from cosmic rays (run 6 is similar to run 5 and thus not shown). For Earth, GCR simulation (run 1) peak NO production occurs near the tropopause (at ~11 km). This is because the GCR energy spectrum (Figure 1) has large values at high energies, which favor secondary electron production deep in the atmosphere (see Figure 2). For the runs with quiescent and active conditions, peak NO production occurs in the mid-stratosphere between (0.03 to 0.04) bar (~20 to 22km). This is because their energy spectrum (Figure 1) is overall weaker than the Earth GCR case, producing secondary electrons at higher atmospheric levels (Figure 2). The extreme "flaring" case in Figure 3 features maximum NO production somewhat lower in the atmosphere at around 0.05 bar.

The rates shown in Figure 3 lead to a concentration of NO for the flaring case in run 4 of $2 \times 10^{14}$ molecules cm$^{-3}$ (~$5 \times 10^{-4}$ volume mixing ratio) in the mid-stratosphere. This value corresponds to within about a factor of two to the peak concentrations calculated by Segura et al. (2010) in this region (see their Figure 8, lower panel). The agreement is good given that we adopt independent methods, that is, Segura et al. use a proxy approach based on X-ray observations and simulate a single large flare, whereas we use a theoretical air shower method as previously described and assume constant, strong background flaring conditions (see also Grenfell et al., 2007[1]).

**3.2 Column Amounts of Biomarkers and Related Compounds**

Observations for Earth suggest that tropospheric NO increases lead to O$_3$ increases via the smog mechanism (Haagen-Smit, 1952), whereas stratospheric NO increases lead to catalytic loss of $O_3$ (Crutzen, 1970). $NO_x$ and $O_3$ interact in different ways in the troposphere, depending upon atmospheric conditions (e.g., UV, $NO_x$, hydrocarbon abundances etc.). A null-cycle involving $NO_x$ recycles $O_3$ as follows:

$$NO+O_3 \rightarrow NO_2+O_2$$
$$NO_2+h\nu \rightarrow NO+O$$
$$O+O_2+M \rightarrow O_3+M$$
*net:* null

The null cycle recycling tropospheric $O_3$ competes with the established "smog cycle," producing $O_3$ in which Volatile Organic Compounds (VOCs) are oxidized by OH in the presence of $NO_x$ and UV. There are three main regimes, firstly, where VOC abundances are at a premium for $O_3$ production ("VOC-controlled"), secondly, where $NO_x$ is at a premium ($NO_x$-controlled), thirdly, in urban areas where $NO_x$ abundances are very high (greater than about $10^{-6}$ vmr). In the latter regime, the removal reaction: $NO_2+OH+M \rightarrow HNO_3+M$, lowers OH, which is a key-player in the smog cycle and, hence, can effectively lower smog $O_3$ production, leading to a significant $O_3$ decrease.

Figure 4 shows column values of some important biomarkers and greenhouse gases. Ozone ($O_3$) columns are reduced by 39% for the quiescent M-dwarf star case (run 3) compared with the Earth GCR case (run 1). The active M-dwarf star case (run 4) features a 63% reduction compared with run 1. This suggests that the stronger cosmic ray-induced NOx effect in run 4 can lead to more $O_3$ loss compared with run 2. The "flaring" M-dwarf star extreme case (run 5) leads to almost complete removal of the $O_3$ atmospheric column, by more than 99% compared with the Earth GCR run, suggesting that the enhanced $NO_x$ photochemistry in this run can act as a major sink for the $O_3$ biomarker.

The main chemical *sink* for $O_3$ switched from the familiar catalytic cycles (Earth GCR run 1) to the single (non-catalytic) titration reaction: $NO+O_3 \rightarrow NO_2+O_2$ for the "flaring" case (run 5). A similar effect is observed on Earth in large cities with very elevated NO abundances (see e.g., Shao et al., 2009). The main $O_3$ *source* in all runs proceeds via: $O+O_2+M \rightarrow O_3+M$, since this is the only reaction that forms $O_3$ in our model. However, the main source of atomic oxygen (O) for the above reaction changes from the familiar photolysis of $O_2$ in the Earth GCR case (run 1) to the photolysis of $NO_2$ via $NO_2+h\nu \rightarrow NO+O$ for the flaring case. Overall, ozone loss for the flaring case was strongly enhanced by CR chemistry.

$O_3$ can be produced or destroyed depending, for example, on the incoming top-of-atmosphere stellar radiative flux and the planetary atmospheric composition. In regions where UV is strong enough to photolyze efficiently, $O_2$ (e.g., in Earth's stratosphere) enhanced NO is expected to destroy $O_3$ via catalytic cycles. In regions where UV is less strong and hydrocarbons and $NO_x$ are in excess of a few tens to hundreds of parts per trillion by volume (pptv), the smog mechanism will operate and enhanced NO can lead to more $O_3$ production. Our results suggest that this effect could be sensitive to changes in the CR input spectrum. For example, Figure 3 shows that the peak NO production rate shifts from the upper troposphere for the "flaring" run, to the upper stratosphere for the quiescent run.

Methane ($CH_4$) columns for the M-dwarf star cases show a large increase compared with the GCR Earth results. An important effect is a decrease in the hydroxyl (OH) radical abundance, an important atmospheric sink for $CH_4$. For example, the OH amount at the planetary surface is reduced by about three orders of magnitude in run 3 compared with the Earth GCR case (run 1). The OH reduction is related to a weakening in the well-known OH-forming reaction: $H_2O+O^1D \rightarrow 2OH$, for the M-dwarf star cases compared with run 1 because the lower UVB output from the M-dwarf star led to a lowering in atmospheric excited oxygen (O$^1$D) abundances[1] and, hence, a slowing in this reaction, since O$^1$D is produced photolytically. In summary, the main effect is that less UVB leads to less OH and, therefore, more CH$_4$ for the M-dwarf star cases. Segura et al. (2005) also discussed CH$_4$ perturbations associated with OH, and as was the case in our work. they also calculated large CH$_4$ abundances of ~4x10$^{-4}$ vmr at the planetary surface for their scenarios of Earth-like planets orbiting M-dwarf stars. Our results suggest that the active case (run 4) and the flaring case (run 5) feature lower CH$_4$ columns than the quiescent case (run 3) (see also section 3.3). Note that the biomarker chloromethane (CH$_3$Cl), like CH$_4$, is emitted at the surface and is removed from the atmosphere mainly via reaction with OH and via photolysis. CH$_3$Cl is, however, much more reactive than CH$_4$, being removed from the atmosphere about 8 times faster. Due to its similar photochemistry, CH$_3$Cl and CH$_4$ responses in the various scenarios are analogous, so CH$_3$Cl is not discussed in detail here.

Nitrous oxide (N$_2$O) columns survive even in the extreme flaring case (run 5). The M-dwarf star radiation input at the top of the model atmosphere is weak in UVB emissions, which weakens the destruction of atmospheric planetary N$_2$O since this molecule is destroyed photolytically (see also UVB text below).

Water (H$_2$O) columns are large for the M-dwarf star cases related in the stratosphere to the large amounts of CH$_4$, which is oxidized to H$_2$O. A damper troposphere is also favored by enhanced greenhouse heating since temperatures here increased by up to 15 K compared with run 1 (see Figure 6).

Table 1 compares above column results from our work (the quiescent M-dwarf star run, i.e.. with weak SCRs), with that of Rauer et al. (2011) who used the same model version and same AD Leo radiation input, but without CRs. Rauer et al. (2011) obtained a ~50% enhanced $O_3$ column for their AD Leo scenario since they omit the CR $NO_x$ source, which catalytically destroys stratospheric $O_3$. Our work calculates a somewhat lower $CH_4$ column than the Rauer et al. study, since in our work the CR $NO_x$ led to faster: $NO+HO_2 \rightarrow OH+NO_2$, that is, more tropospheric OH, which destroyed more $CH_4$. Finally, the lower $O_3$ column (hence higher UV radiation) in our work, for example, led to about 17% lower $N_2O$ in the column compared with the findings of Rauer et al. since this species is destroyed photolytically by UV radiation.

---

[1] The notation following the species name is called the Term Symbol and takes the form: $^{M(2S+1)}$L where M is the multiplicity, S is the spin quantum number and L is the total orbital momentum quantum number. The Term Symbol conveys information about the electronic state of the chemical species considered.

UVB (280-315 nm) radiation at the planetary surface plays a critical role for biological organisms. The amount of UVB reaching the surface is determined in our model by the TOA input stellar spectrum, the amount of gaseous $O_3$ absorption, and by Rayleigh scattering of atmospheric molecules. The values listed in Table 2 suggest that, for Earth (run 1), surface UVB in our study compares reasonably well with that of the Segura et al. (2010) study but is somewhat higher than the Segura et al. (2005) study and Earth observations, although these observations are not well determined globally. Run 2 features lower surface UVB by about three orders of magnitude compared with run 1, since M-dwarfs are weak emitters of UVB radiation.

For the flaring run (with very low $O_3$ absorption), Table 2 suggests that the Rayleigh scattering component (arising from $O_2$, $N_2$ and $CO_2$) becomes important for surface UVB – this is apparent because, for the comparison case (run 6) with no Rayleigh scattering, the UVB radiation is no longer shielded by the atmosphere. Future work will investigate the influence of Rayleigh scattering upon shielding of UVB and related surface habitability. Also, note that $O_3$ is the only gas-absorber in the chemistry module that contributes to UVB absorption through the atmosphere. Future work will test the effect of including trace-species that absorb in the UVB, although the effect, at least for Earth, is expected to be small.

**3.3 Atmospheric Profiles of Biomarkers and Related Compounds**

The $O_3$ profile (Figure 5a) shows that the $O_3$ destruction in the flaring case (run 5) is severe and takes place over the entire atmospheric column. It also implies that a weaker $O_3$ layer in the active M-dwarf star (run 4) and flaring scenario (run 5) led to some so-called "self-healing" (i.e., a reduction in $O_3$ on the upper atmospheric levels led to more UV in the Herzberg wavelength region penetrating to the underlying regions, where faster $O_2$ photolysis favored enhanced $O_3$ production). Figure 5a features a local maximum in the $O_3$ abundance in the upper troposphere for these two cases.

With regard to the $CH_4$ profile (Figure 5b), the quiescent M-dwarf star run, in comparison with the Earth GCR (run 1), features more surface $CH_4$ by a factor of ~207. A large $CH_4$ greenhouse effect was also noted by Segura et al. (2005) as already discussed above. A goal of our work is to study $CH_4$ responses with gradually increasing SCR $NO_x$ sources. In Earth's atmosphere, $CH_4$ removal is mostly controlled by OH, a member of the $HO_x$ family - defined here as the sum of (OH+$HO_2$). The two family members are subject to mutual interconversion, which involves chemical reactions quickly destroying OH to form $HO_2$ and vice-versa. The two species quickly attain a dynamical equilibrium with a particular concentration partitioning (depending on, for example, altitude, location, etc.) where chemical loss and production rates are in steady-state. Introducing more nitrogen oxides (e.g., induced by stronger SCR activity) has the pivotal effect of driving the $HO_x$ partitioning away from $HO_2$ into OH via one of the quickest-known reactions in the Earth's lower atmosphere:

$$NO+HO_2 \rightarrow NO_2+OH$$

-which forms OH and, therefore, leads to $CH_4$ reduction with increasing $NO_x$ sources. Consistent with this reasoning, $CH_4$ abundances in Figure 5b are reduced by about 43% for the active star scenario compared with the quiescent star scenario. For the flaring scenario (run 5), $NO_x$ sources increase drastically, leading to an enhancement in OH based on reaction (5) and, hence, a decrease in $CH_4$.

The $N_2O$ profile (Figure 5c) featured about (3 to 8)x$10^{-7}$ surface vmr for runs 2 to 5, that is, more than twice as much as on Earth. $N_2O$ abundances on Earth are sensitive to UV, which is its main atmospheric sink. On comparing the UVB fluxes from the M-dwarf star runs with the Earth observations (Table 1), there are two main effects. First, the M-dwarf stellar spectrum applied at the top of the atmosphere is weak in the UVB range (typically a factor of 50 to 100 lower than the Solar value). Second, clearly the amount of UVB reaching the model surface will depend on the column abundance of strong UVB absorbers such as $O_3$ (which is very low, e.g., for the flaring case as already mentioned) and on the Rayleigh scattering. For example, although the flaring case featured almost no $O_3$ layer, its surface nevertheless received a dosage of UVB about 7 times *less* than is observed for Earth (Table 2). This was due to (a) the weaker stellar UVB radiation and (b) shielding by Rayleigh scattering in the model, which protects the planet's surface in the flaring case even when the $O_3$ layer is removed. This effect also helps the $N_2O$ biomarker to survive. Table 2 suggests that the effect of changing the stellar emissions on surface UVB is stronger than the effect related to changes in atmospheric composition by SCRs.

For the $H_2O$ profile (Figure 5d), abundances rose from about 8 parts per million by volume (ppmv) in Earth's mid-stratosphere (run 1) up to about 100 ppmv in this region for, example, the flaring case (run 5). The damp stratosphere was favored by high $CH_4$ levels, since $CH_4$ oxidation is a major source of $H_2O$. In the free troposphere, warming from, for example, the $CH_4$ greenhouse led to an increase in the atmospheric water due to stronger evaporation.

### 3.4 Atmospheric Temperature Profiles

Figure 6 shows temperature profiles. The M-dwarf star scenarios display a cool middle atmosphere (see also Segura et al., 2005; Rauer et al., 2011) with a notably weak altitude dependence compared with the Earth GCR case (run 1). This arises, for example, due to weak UVB radiation output of the stars and hence involves weaker UVB heating. A smaller effect also arises from the more suppressed $O_3$ abundances in the M-dwarf star scenarios since this species on Earth is a major stratospheric heater and is responsible for the strong temperature increase with height in the Earth GCR run. In addition, the enhanced $H_2O$ and $CH_4$ in the M-dwarf star scenarios also played a significant role for the stratospheric radiative heating budget. In summary, changing the SCRs had only a relatively small impact on the atmospheric temperature profiles in the M-dwarf star cases.

### 3.5 Thermal Emission Spectra

Figure 7 illustrates the theoretical emission spectrum of an Earth-like planet for the quiescent and the flaring M-dwarf star cases (runs 3, 5). Results are shown at a spectral resolution, $R_s =100$, over the wavelength range from (2-5) microns for the upper panel and from (5-20) microns for the lower panel. The Figure suggests that the fundamental $O_3$ band at 9.6 microns disappears for the flaring case since $O_3$ is efficiently destroyed by stellar cosmic-ray induced photochemical effects as already discussed. Also, the prominent $CO_2$ bands, for example, at 15 microns are enhanced for the flaring case in Figure 7. Although $CO_2$ abundances are kept constant in all runs (to 355 ppmv, see section 2.1), the 15 micron band for the flaring case arises mainly at an atmospheric height, which has a colder brightness temperature and results in a deeper 15 micron spectral band than for, for example, the quiescent M-dwarf star case (run 3). Finally, Figure 7 suggests stronger absorption in the temperature-sensitive 6.3 micron $H_2O$ vibrational-rotation band for the flaring case (run 5). The brightness temperature of this band for the flaring case was 250 K, that is, about 15 K colder than for run 3, and occurred in the upper troposphere at about 0.3 bar. Here, the water abundances for the flaring case were about 50% lower than the quiescent case, which suggests that the brightness temperature effect and not the change in water abundance was driving the water band enhancement for the flaring run spectrum. Figure 7 also suggests an enhancement in the 11.3 micron band of nitric acid ($HNO_3$), a species formed by oxidation of $NO_x$ by OH, which in our results could represent a spectroscopic indicator for high flaring activity of the star.

## 4. Discussion and Modeling Caveats

Our main result is that $O_3$ is removed for strong SCR activity conditions. We now place this result in context by discussing our assumptions and possible responses for Earth-like planets that are difficult to capture in our current model. As mentioned, we assume the upper-limit case where the planet has no magnetic field. Whether our considered planets are indeed tidally locked and the effect on their magnetospheres is currently under debate in the literature and is beyond the scope of the present study.

$O_3$ 3D transport effects were discussed earlier (section 2.1) for Earth. For exoplanets, we now briefly mention some potentially important effects. Critical for the M-dwarf cases is the rate of transport of $O_3$ (produced on the dayside) to the planet's nightside. This will depend, for example, on the atmospheric mass, the day-night temperature gradient, and the planet's rotation rate; hence, its Coriolis force. Seasonal $O_3$ effects depend on planetary obliquity, eccentricity, or inclination. If the exo-winter is longer than the $O_3$ photochemical timescales, which are a few weeks on Earth, then $O_3$ would build up in the exo-winter. Latitudinal transport of $O_3$ via the "meridional circulation" will be strengthened by exo-orography because this favors atmospheric wave activity, which drives the circulation. Faster planetary rotation will slow the meridional circulation via a strengthened Coriolis force, which disrupts equator-to-pole transport. Since the polar "$O_3$ hole" on Earth is mainly associated with chlorine released from industrial chlorofluorocarbons (CFCs), planets with strong emissions of naturally occurring chlorine-containing biomass (e.g., $CH_3Cl$ emitted from seaweed) may feature their own "natural" polar $O_3$ holes. The altitude dependence of $O_3$ depends, for example, on the trade-off between (Herzberg-region) UV radiation, which can photolyze $O_2$ (favouring $O_3$ at higher altitudes) on the one hand, and the abundance of $O_2$ (favoring $O_3$ at lower altitudes) on the other hand. Exoplanets that are more transparent to such UV or/and with thick Earth-like atmospheres may favor a downward shift in the altitude where $O_2$ photolysis peaks and, hence, since this favors $O_3$ formation, to a downward shift in their $O_3$ maximum. $O_3$ loss via catalytic cycles may also play a role, for example, exoplanets with high lightning activity (hence, high $NO_x$) or warm, moist waterworlds (with high $HO_x$).

Transport and mixing are important factors for flare-related CR-induced $O_3$ loss in Earth-like exoplanet scenarios. Critical is how much enhanced $NO_x$ (from CRs) is transported to the planet's $O_3$-layer. On Earth, high NO formed above the mesosphere (where it is quickly destroyed in lit regions) is transported downwards in the dark polar vortex, until it reaches the (lower) stratosphere. At these altitudes, the barrier to mixing at the vortex edge is weaker, and air can escape the vortex to reach sunlit regions at mid-latitudes, where considerable $O_3$ loss occurs.

Eddy-diffusion coefficients (K) are a rather straightforward means in 1D models to represent the physics of 3D atmospheric mixing. Patra and Lal (1997) provided a review and suggested global mean K values similar to our work. The quoted range for K in their paper is, however, quite large - more than a factor 10 depending on latitude and season. As discussed earlier, a test case (run 6) with reduced Eddy mixing still displayed strong $O_3$ removal of >99%, which suggests that the uncertainties in Eddy mixing do not affect our main conclusions.

Atmospheric mass is another critical quantity that controls, for example, the altitude at which CR shower events peak. We assumed 1 bar surface pressure in all our scenarios. This is, however, currently unconstrained for Super-Earths and depends on, for example, delivery, outgassing, and escape processes. Our results for the flaring scenario indicate peak NO production at about 0.1 bar, but we note that a wide range of atmospheric masses are possible. For example, surface pressures of the terrestrial planets in the Solar System range from $10^{-15}$ bar on Mercury up to ~92 bar on Venus.

## 5. Summary

The scenarios considered the effect of CRs on $N_2$-$O_2$ atmospheres with $P_{surface}$=1bar, with mean (day-night averaged) insolation and for a planet (orbiting in the HZ of an M-dwarf star) with similar surface gravity and Eddy mixing as Earth. The atmospheric column model used is applied to these rather Earth-like conditions to remain within its range of validity. A future model version will be applied over a much wider range of composition, temperature, and pressure conditions.

Our theoretical spectra calculations suggest that biomarker spectral signals of planets orbiting in the HZ of very active flaring M-dwarf stars may show a strong suppression of the $O_3$ fundamental band (9.6 microns), a deeper $CO_2$ band (15 microns), and an enhanced $H_2O$ vibrational-rotational band (6.3 microns) compared with less active M-dwarf stars. Our work therefore suggests that planets orbiting the HZ of strongly flaring stars may feature low $O_3$ spectroscopic signals. About 12% of M-dwarfs within 7 parsecs have stronger flaring indices than AD Leo, based on ($L_{x-ray}$/$L_{bolometric}$) measurements by Fleming et al. (1995). Improved catalogues of flaring activity for M-stars (as discussed, e.g., in Reiners et al. 2012) as well as improved stellar spectra in the UV and EUV, which could strongly affect calculated $O_3$ abundances, are desirable.

Surface UVB, a critical quantity for habitability, is low for the M-dwarf star cases compared with Earth observations (Table 2) since, for example, the UVB emission of M-dwarf stars is weak. Also, the level of UVB atmospheric protection by absorption depends, for example, on the planetary $O_3$ column. Also, important Rayleigh scatterers in the model affecting UV are $N_2$, and $O_2$, whose modeled concentrations are fixed. Biomarkers $O_3$ and $N_2O$ are generally quite resilient to $NO_x$-induced photochemical effects from SCRs for Earth-like planets orbiting in the MHZ. $O_3$, however, may not survive extreme, steady, flaring conditions.

**Acknowledgements**
This research has been supported by the Helmholtz association through the research alliance "Planetary Evolution and Life". We are grateful to Prof. James Kasting for providing original code and for useful discussion. P.v.P. acknowledges support from the European Research Council (Starting Grant 209622: E3ARTHs).

**References**

Amsler, C., Doser, M., Antonelli, M., Asner, D. M., Babu, K. S., and 173 others, Passage of particles through matter, Physics Lett.. B667, 1, 2008.

Audard, M., Guedel, M., Kraje, J., and Kashyap, V. L., Extreme ultra-violet flare activity in late-type stars, ApJ., 541, 396-409, 2000.

Bichsel, H., Groom, D. E., and Klein S. R., Passage of particles through matter, Chapter 27 in S. Eidelman et al., Rev. Particle Physics, Phys. Lett. B592, 2005.

Brasseur, G. P., Orlando, J. J., and Tyndall, G. S. (eds), Atmospheric chemistry and global change, Oxford University Press, 1999.

Chapman, S., A theory of upper atmosphere ozone, Mem. Roy. Soc., 3, 103-125 (1930).

Crutzen, P. J., Influence of Nitrogen Oxides on Atmospheric Ozone Content, Quart. J. Roy. Met. Soc., 96, 320-325, 1970.

Dartnell, L. R., Ionizing radiation and life, Astrobiology, 11, 551-582, 2011.

Dwyer, J. R., The initiation of lightning by runaway air breakdown, Geophys. Res. Lett, 32, L20808, 2005.

Ejzak, L. M., Melott, A. L., Medvedev, M. V., and Thomas, B. C., Terrestrial consequences of spectral and temporal variability in ionizing photon events, Astrophys. J., 654, 373-384, 2007.

Feynman, J. G., and Gabriel, S., The effects of high energy particles on planetary missions, Feynman & Gabriel (Eds.), Interplanetary Particle Env., proceedings of a conference, JPL publication 88-28, pp. 35-47, 15th April 1988.

Feynman, J., Spitale, G., Wang, J., and Gabriel, S., Interplanetary proton fluence model, J. Geophys. Res., 98, 13281-13294, 1993.

Fleming, T A., Schmitt, J. H. M. M., Giampa, M., Correlations of coronal X-ray emission with activity, mass and age of the nearby K and M dwarfs, Astrophys. J., 450, 401-410, 1995.

Gaisser, T.K., and Hillas, A. M., Reliability of the Method of Constant Intensity Cuts for Reconstructing the Average Development of Vertical Showers, Proc. 15th Int. Conf. RC (Plavdiv, 1977), 8-353, 1977.

Gueymard, C. A., The sun's total and spectral irradiance for solar energy applications and solar radiation models, Solar Energy, 76, 423-453, 2004.

Gloeckler, G., Fisk, L. A., Geiss, J., Hill, M. E., Hamilton, D. C., Decker, R. B., and Krimigis, S. M., Composition of interstellar neutrals and origin of anomalous cosmic rays, Spa. Sci. Rev., doi 10.1007/ss11214-008-9482-5, 2009.

Grenfell, J. L., Grießmeier, Patzer, B., Rauer, H., Segura, A., Stadelmann, A., Stracke, B., Titz, R., and von Paris, P., Biomarker Response to Galactic Cosmic Ray-Induced $NO_x$ and the Methane Greenhouse Effect in the Atmosphere of an Earthlike Planet Orbiting an M-Dwarf Star, Astrobiol., 7 (1), 208-221, 2007[1].

Grenfell, J. L., Stracke, B., von Paris, P., Patzer, B., Titz, R., Segura, A., and Rauer, H., The Response of Atmospheric Chemistry on Earthlike Planets around F, G, and K stars to Small Variations in Orbital Distance, Plan. Spa. Sci. Special Issue on Habitable Zones, Plan. Spa. Sci., 55, 661-671, 2007[2].

Grenfell, J. L., Gebauer, S., von Paris, P., Godolt, M., Hedelt, P., Patzer, A. B. C., Stracke, B., and Rauer, H., Sensitivity of biomarkers to changes in chemical emissions in the Earth's Proterozoic atmosphere, Icarus, 211, 1, 81-88, 2011.

Grießmeier, J.-M., A. Stadelmann, U. Motschmann, N. K. Belisheva, H. Lammer, and H. K. Biernat, Cosmic ray impact on Extrasolar Earth-like planets in close-in habitable zones, Astrobiol., 5(5), 587-603, 2005.

Grießmeier, J.-M., Stadelmann., A., Grenfell, J. L., Lammer, H., and Motschmann, U., On the protection of extrasolar Earth-like planets around K/M stars against galactic cosmic rays, Icarus, 199, 526-535, 2009.

Grießmeier, J.-M., Khodachenko, M., Lammer, H., Grenfell, J. L., Stadelmann., A., and Motschmann, U., Stellar activity and magnetic shielding, in Solar and Stellar Variability: Impact on Earth and Planets, IAU Symposium, 264 (Editors) Kosovichev, A. G., Andrei, A. H., and Rozelot, J.-P., Proceedings of the Astronomical Union, 385-394, 2010.

Gopalswamy, N., Coronal mass ejections of solar cycle 23, J. Astrophys. Astron., 27, 243-254, 2006.

Güdel, M., M. Audard, V. L. Kashyap, J. J. Drake, and E. F. guinan, Are coronae of magnetically active stars heated by flares? II. Extreme ultraviolet and X-ray flare statistics and the differential emission measure distribution, ApJ, 582, 423-442, 2003.

Gueymard, C., The sun's total and spectral irradiance for solar energy applications and solar radiation models, Solar Energy, 76, 423-453, 2004.

Haagen-Smit, A. J., Chemistry and physiology of Los Angeles Smog, Ind. Eng. and Chem., 44, 1342-1346, 1952.

Hamilton, D. C., The radial transport of energetic solar flare particles from 1 to 6 AU, J. Geophys. Res., 82, 2157-2169, 1977.

Hamilton, D.C., G.M. Mason, and F.B. McDonald, The radial dependence of the peak flux and fluence in solar energetic particle events, Proc. 21st Intl. Cosmic Ray Conf., (Adelaide), 5, 237-240, 1990.

Hauschildt, P. H., Allard, F., Ferguson, J., Baron, E., and Alexander, D. R., The NEXGEN model atmosphere grid. II. Spherically symmetric model atmospheres for giant stars with effective temperatures between 3000 and 6800K, Astrophys. J., 525, 871-880, 1999.

Houdebine, E. R., Foing, B., and Rodono, M., Dynamics of flares on late-type dMe stars, Astron. Astrophys., 238, 249-255, 1990.

Jackman, C. H., Frederick, J. E., and Stolarski, R. S., Production of odd nitrogen in the stratosphere and mesosphere: an intercomparison of source strengths, J. Geophys. Res., 85, doi: 10.1029/0JGREA000085000C12007495000001.

Jackman, C. H., Fleming, E., and Vitt, F., Influence of extremely large solar proton events in a changing stratosphere, J. Geophys. Res., 11,659-11,670, 2000.

Jackman, C. H., DeLand, M. T., Labow, G. J., Fleming, E. L., Weisenstein, D. K., Ko, M. K. W., Sinnhuber, M., and Russell, J. M., Neutral atmospheric influences of the solar proton events in October-November 2003, J. Geophys. Res., 110, doi: 10.1029/2004JA010888, 2005.

Jackman, C. H., Marsh, D. R., Vitt, F. M., Roble, R. G., Randall, C. E., Bernath, P. F., Funke, B., Lopez-Puertez, M., Versick, S., Stiller, G. P., Tylka, A. J., and Fleming, E. L., Atmos. Chem. Phys., 11, 6153-6166, 2011.

Jet Propulsion Laboratory Publication 02-25, Chemical Kinetics and Photochemical Data for use in Atmospheric Studies, Evaluation No. 14, February 2003.

Khodachenko, M. L., I. Ribas, H. Lammer, J. -M. Grießmeier, M. Leitner, F. Selsis, C. Eiroa, A. Hanslmeier, H. K. Biernat, C. J. Farrugia, and H. O. Rucker, Coronal Mass Ejection (CME) activity of low mass M-stars as an important factor for the habitability of terrestrial planets. I: CME impact on expected magnetospheres of Earth-like exoplanets in close-in habitable zones, Astrobiol., 7(1), 167-184, 2007.

Kuznetsov, N. V., R. A. Nymmik, and M. I. Panasyuk, The balance between fluxes of galactic cosmic rays and solar energetic particles, depending on solar activity, 27[th] International Cosmic Ray Conference (ICRC), 3193, 2001.

Kuznetsov, N. V., R. A. Nymmik, and M. I. Panasyuk, Models of solar energetic particle fluxes: The main requirements and the development prospects, Adv. Spa. Res. 36, 2003–2011, 2005.

Lammer, H, Lichtenegger, H. I. M., Kulikov, Y., Grießmeier, J. –M., Terada, N., Erkaev, N. V., Biernat, H. K., Khodachenko, M. L., Ribas, I., Penz, T., and Selsis, F., Coronal Mass Ejection (CME) activity of low mass M-stars as an important factor for the habitability of terrestrial exoplanets. II. CME-induced ion pick-up of earth-like exoplanets in close-in habitable zones, Astrobiol., 7, 1, 185-207, 2007.

Lammer, H., Kasting, J. F., Chassefiere, E., Johnson, R. E., Kulikov, Y. N., and Tian, F., Atmospheric escape and evolution of terrestrial planets satellites, Spac. Sci., Rev., 139, 399-436, 2008.

Lario, D., R. G. Marsden, T. R. Sanderson, M. Maksimovic, B., Sanahuja, A., Balogh, R. J., Forsyth, R. P., Lin, and J. T. Gosling, Energetic proton observations at 1 and 5 AU I: January-September 1997, J. Geophys. Res., 105, 18235-18250, 2000.

Lario, D., M.-B. Kallenrode, R.B. Decker, E.C. Roelof, S.M. Krimigis, A. Aran and B. Sanahuja, Radial and Longitudinal Dependence of Solar 4-13 MeV and 27-37 MeV Proton Peak Intensities and Fluences: Helios and IMP 8 Observations, Astrophys. J., 653, 1531-1544, 2006.

Leggett, S. K., Allard, F., Berriman, G., Dahn, C. C., and Hauschildt, P. H., Infrared spectra of low-mass stars: towards a temperature scale for red dwarfs, Astrophys. J. Supp., 104, 117, 1996

Manabe, S., and R. T., Wetherald, Thermal equilibrium of the atmosphere with a given distribution of relative humidity, J. Atm. Sci., 24(3), 241-259, 1967.

Melott, A. L. Thomas, B. C., Astrophysical ionizing radiation and the Earth: a brief review and census of intermittent intense sources, Astrobiol., 11, 343-361, 2011.

Melsheimer, C., Verdes C., Buehler, S. A., Emde, C., Eriksson, P., Feist, D. G., Ichzawa, S., John, V. O., Kasai, Y., Kopp, G., Koulev, N., Kuhn, T., Lemke, O., Ochiai, S., Schreier, F., Sreerekha, T. H., Suzuki, M., Takahashi, C., Tsujimaru, S., and Urban, J., Intercomparison of general purpose clear sky atmospheric radiative transfer models for the millimeter/submillimeter spectral range, Radio Science, 40, RS1007, 10.1029/2004RSS003110, 2005.

Mlawer, E. J., Taubman, S. J., Brown, P. D., Iacono, M. J., and Clough, S. A., Radiative transfer model for inhomogeneous atmospheres: RRTM, a validated correlated-k model of the longwave, J. Geophys. Res., 102, 16,663-16,682, 1997.

Nicolet, M., On the production of nitric oxide by cosmic rays in the mesosphere and stratosphere, Plan. Spa. Sci., 23, 637-649, 1975.

Particle Data Group (PDG) Review of Particle Physics, Physical Review (D), 66, 010001, 2002.

Patra, P. K., Lal., S., Variability of eddy diffusivity in the stratosphere deduced from vertical distributions of $N_2O$ and CFC-12, J. Atm. Sci. Sol. Terr. Phys., 59, 1149-1157, 2997.

Pavlov,, A., A. and Kasting, J. F., Mass-independent fractionation of sulfur isotopes in Archean sediments: strong evidence for an anoxic Archean atmosphere. Astrobiol., 2, 27-41, 2002.

Pettersen, B. R., Hawley, S. L., A spectroscopic survey of red dwarf flare stars, Astron. Astrophys., 217, 187-200, 1989.

Porter, H. S., Jackman, C. H., Green, A. E. S., Efficiencies for production of atomic nitrogen and oxygen by relativistic proton impact in air, J. Chem. Phys., 65, 154, 1976.

Rauer, H., Gebauer, S., v. Paris, P., Cabrera, J., Godolt, M., Grenfell, J. L., Belu, A., Selsis, F,. Hedelt, P., and Schreier, F., Biomarkers in Super-Earth Atmospheres, I. Spectral appearance of Super-Earths around M-dwarfs, A&A, 529 (A8), doi: 10.1051/0004-6361/201014368, 2011.

Reiners, A., Joshi, N., and Goldman, B., A catalogue of rotation and activity in early M-stars, AJ, accepted, 2012.

Rothman, L. S., Jacquemart, D., Barbe, A., Chris Benner, D., Birk., M. et al., The HITRAN 2004 molecular spectroscopic database, J. Quant. Spec. and Rad. Transfer, 96, 139-204, 2005.

Ruzmaikin, A., G. Li, and G. Zank, The radial dependence of solar energetic particle fluxes. Proc. Conference Solar Wind 11-

SOHO 16, Connecting Sun and Heliosphere, ESA SP-592, 441–444, 2005.

Scalo, J., Segura, A., Fridlund, M., Ribas, I,. Odert, P., Leitzinger, M., Kulikov, Y. N., Grenfell, J. L., Rauer, H., Kaltenegger, L., Khodachenko, M. L., Selsis, F., Eiora, C., Kasting, J.,  and Lammer, H., M-type dwarf stars and their relevance for terrestrial planet finding missions, Astrobiol., 7 (1), 85-166, 2007.

Scherer, K., Fichtner, H., and Stawicki, O., Shielded by the wind: the influence of the interstellar medium on the environment of Earth, J. Atm. Sol. Terr. Phys., 64, (7), 795-804, 2002.

Schreier, F., and Schimpf, B., A new efficient line-by-line code for high resolution atmospheric radiation computations incl. derivatives. In Smith, W. L., and Timofeyev, Y., editors, IRS, 2000: Current problems in atmospheric radiation, 381-384. A. Deepak Publishing, 2001.

Segura, A., Krelove, K., Kasting, J. F., Sommerlatt, D., Meadows, V., Crisp, D., Cohen, M., and Mlawer, E., Ozone concentrations and ultraviolet fluxes on earth-like planets around other stars, Astrobiol., 3, 689-708, 2003.

Segura, A., Kasting, J. F., Meadows, V., Cohen, M., Scalo, J., Crisp, D., Butler, R. A. H., Tinetti, G., Biosignatures from Earth-like planets around M-stars, Astrobiol., 5, 706-725, 2005.

Segura, A., Walkowicz, L. M., Meadows, V., Kasting, J., F. and Hawley, S., The effect of a strong flare on the atmospheric chemistry of an earth-like planet orbiting an M-star, Astrobiol. 10 (7), 751-771, 2010.

Seo, E. S., Ormes, J. F., Streitmatter, R. E., Stochaj, S. J., Jones, W. V., Stephens, S. A., and Bowen, T., Measurement of cosmic ray proton and helium spectra during the 1987 solar minimum, Astrophys, J., 378, 763-772, 1991.

Shao, M., Lu, S. H., Liu, Y., Xie, X., Chang, C. C., Hunag, S., and Chen, Z. M., Volatile organic compounds measured in summer in Beijing and their role in ground-level ozone formation, J. Geophys. Res., 114, doi: 10.1029/2008JD010863, 2009.

Smart D. F., and Shea, M. A., A review of solar proton events during the 22nd solar cycle, Adv. Sp. Res., 30 (4), 1033-1044, 2002.

Svensmark, H., and Christensen, E. F., Variation of cosmic ray flux and global cloud coverage – a missing link in solar-climate relationships, J. atmos. sol.-terr. Phys. 59, 1225-1232, 1997.

Tarduno, J., Cottrell, R. D., Watkeys, M. K., Hofmann, A., Doubrovine, P. V., Mamajek, E. E., Liu, D., Sibeck, D. G., Neukirch, L. P., and Usui, Y., Geodynamo, Solar Wind, and Magnetopause 3.4 to 3.45 Billion Years Ago, Science, 327, 1238-1240, 2010.

Thomas, B. C., Jackman, C. H., Melott, A. L., Laird, C. M., Stolarski, R. S., Gehrels, N., Cannizo, J. K., and Hogan, D. P., Terrestrial ozone delpletion due to a Milky Way Gamma Ray Burst,  Astrophys. J., 622, L153-L156, 2005[a].

Thomas, B. C., Melott, A. L., Jackman, C. H., Laird, C. M., Medvedev, M. V., et al., Gamma ray bursts and the Earth: Exploration of atmospheric, biological, climate and biogeochemical effects, Astrophys. J., 634, 509-533, 2005[b].

Vardavas, I, M., and Carver, J. H., Solar and terrestrial parameterizations for radiative-convective models, Plan. Spa. Sci., 32, 1307-1325, 1984.

von Clarmann, T., Höpfner, M., Funke, B., López-Puertas, M., Dudhia, A., Jay, V., Schreier, F., Ridolfi, M., Cecherini, S., Kerridge, B. J., Reburn, J., and Siddans, R., Modelling of atmospheric mid-infrared radiative transfer: the AMIL2DA algorithm intercomparison experiment, J. Quant. Spec. and Radiat. Transfer, 78, 381-407, 2002.
von Paris, P., The atmospheres of Superearths, Technische Universität Berlin,  Doctoral Thesis, 2010.

Wang, P., Li, Z., Cihlar, J., Wardle, D. I., and Kerr, J., Validation of an UV inversion algorithm using satellite and surface measurements, J. Geophys. Res., 105, D4, 5037-5048, 2000.

World Meteorological Organization (WMO) Scientific Assessment of Ozone Depletion, Report Number 37, 1994.

Ziegler, J. F., Terrestrial cosmic rays, IBM. J. Res. Develop. 40, 1, 19-39, 1996.

**Tables**

Table 1: Comparison of atmospheric column (Dobson Units, DU) of biomarkers and related species from our work (run 3) with the AD Leo scenario from Rauer et al. (2011).

| Species | Run 3 (this work) Weak Cosmic Rays (DU) | Rauer et al. (2011) Without Cosmic Rays (DU) |
|---|---|---|
| Ozone | 183 | 270 |
| Methane | $2.73 \times 10^5$ | $3.48 \times 10^5$ |
| Nitrous Oxide | 551 | 666 |

Table 2: Global Mean UVB at the planetary surface and its ratio between TOA and the planetary surface for the model runs with SCRs and an additional test run without Rayleigh scattering. Output corresponds to (280-315)nm. Reference values are based on Earth measurements of the Total Ozone Monitoring Satellite (TOMS) (Wang et al. 2000), as well as calculated data from Segura et al. (2005, 2010).

| Scenario | UVB (surface) [$Wm^{-2}$] | Ratio UVB (TOA/surface) | Reference |
|---|---|---|---|
| Earth GCR | 2.3[#] | 8.0 | This work Run 1 |
| Earth no GCR | 1.3 | 11.7 | Segura et al. (2005) |
| Earth no GCR | ~2.6 | 6.8 | Segura et al. (2010) |
| Earth Obs. | 1.5* | | Wang et al. (2000) |
| M-star spectrum GCRs | $2.0 \times 10^{-2}$ | 10.0 | This work Run 2 |
| Quiescent M-dwarf SCRs | $5.8 \times 10^{-3}$ $2.8 \times 10^{-2}$ | 6.0 7.4 | Segura et al. (2005) This work Run 3 |
| Active M-dwarf SCRs | $3.9 \times 10^{-2}$ | 5.5 | Run 4 |
| Flaring M-Dwarf SCRs | 0.17 | 1.9 | Run 5 |
| Flaring No Rayleigh | 0.24 | 1.0 | Run 7 |

# Solar mean conditions based on Gueymard (2004)
*For (1992-94) based on 6 ground stations and the Total Ozone Monitoring Satellite (TOMS), Wang et al. (2000). Global mean value is uncertain.

**Figures**

Figure 1: Cosmic Ray Proton Spectra normalised to 0.153 AU and the average GCR spectra for Earth at 1AU (solid line).

Figure 2: Number of secondary electrons ($S_e$) produced relative to the shower maximum ($Nel_{max}$), where $Nel_{max}=3$ for all energies. Data are shown for run 1 which is the case for the Earth orbiting the Sun including the effect of GCRs ("Earth GCR run").

Figure 3: Nitrogen monoxide (NO) production rate varying throughout the atmosphere.

Figure 4: Column values of biomarkers and greenhouse gases in Dobson Units (DU) (1DU= $2.69 \times 10^{16}$ molecules cm$^{-2}$).

Figure 5: Concentration profiles for atmospheric biomarkers and related species (logarithm of the vmr) for (a) ozone (b) methane (c) nitrous oxide and (d) water.

Figure 6: Atmospheric temperature (K) profiles.

Figure 7: Thermal emission spectra ($R_s$=100) for run 3 (quiescent M-dwarf star) and run 5 (steadily-flaring M-dwarf star).

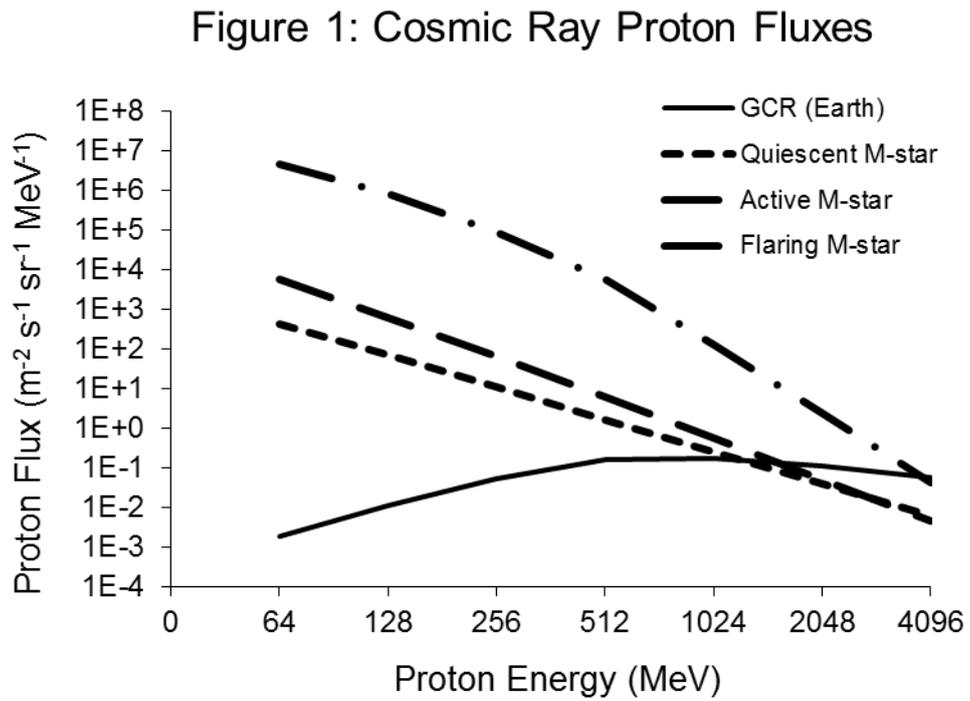

**Figure 2: Number of Secondary Electrons produced relative to Shower Maximum**

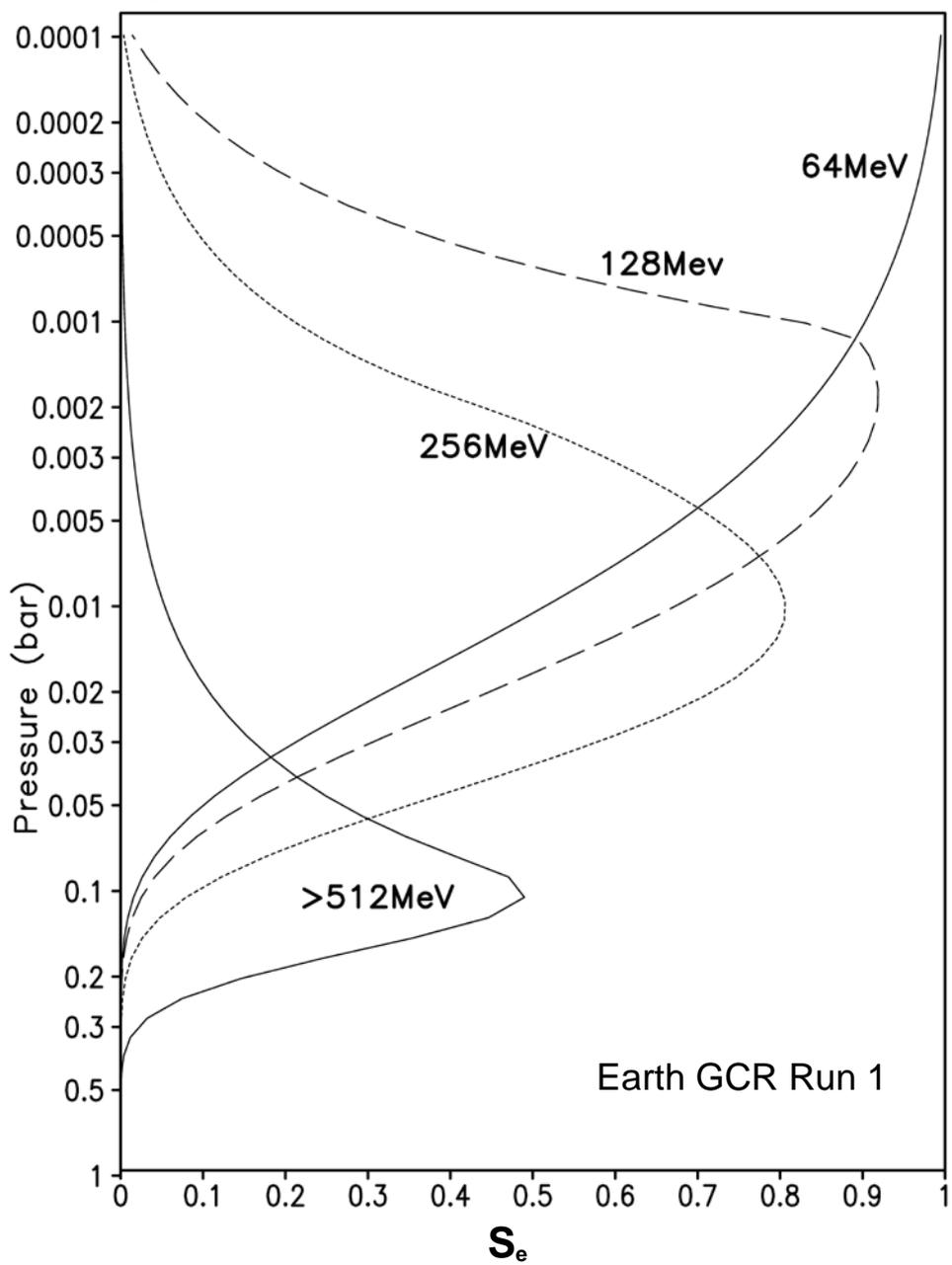

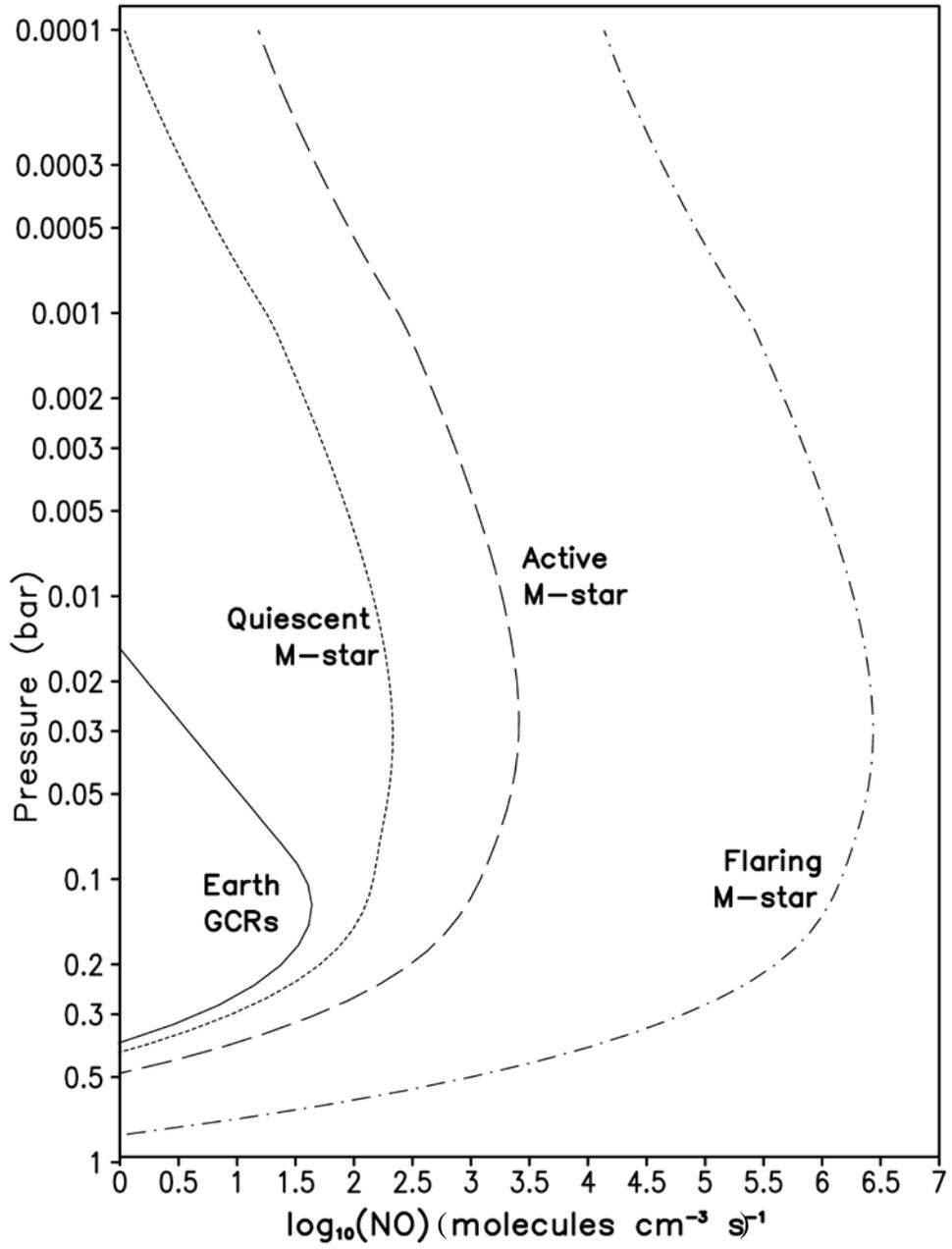

Figure 3: NO production from Cosmic Rays

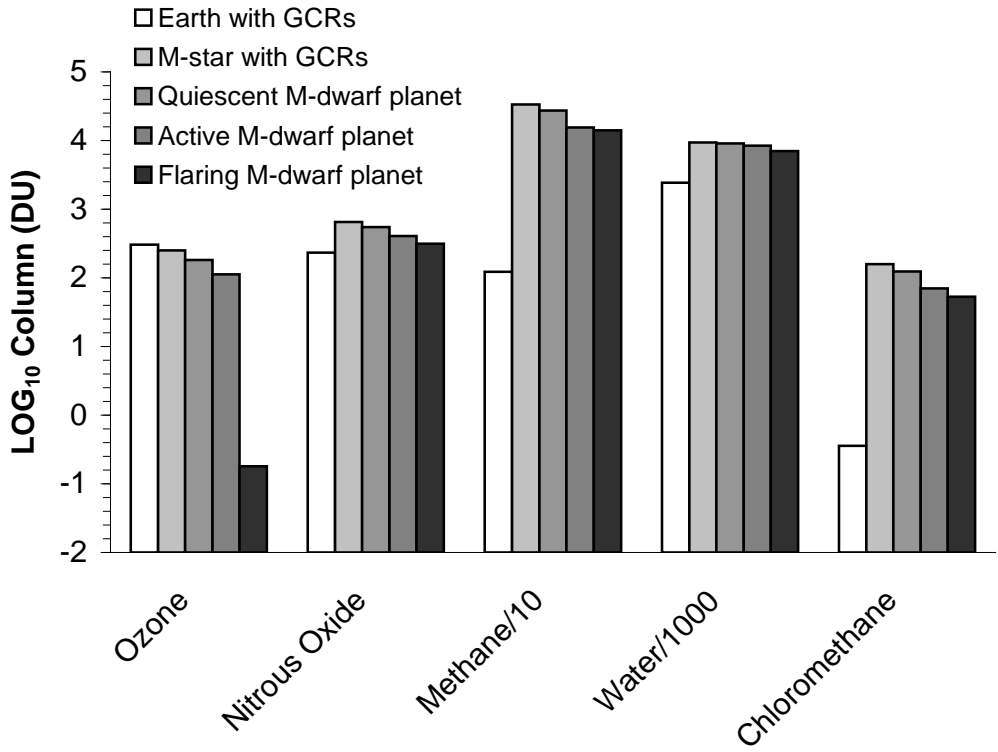

**Figure 4: Biomarkers and Greenhouse Gases**

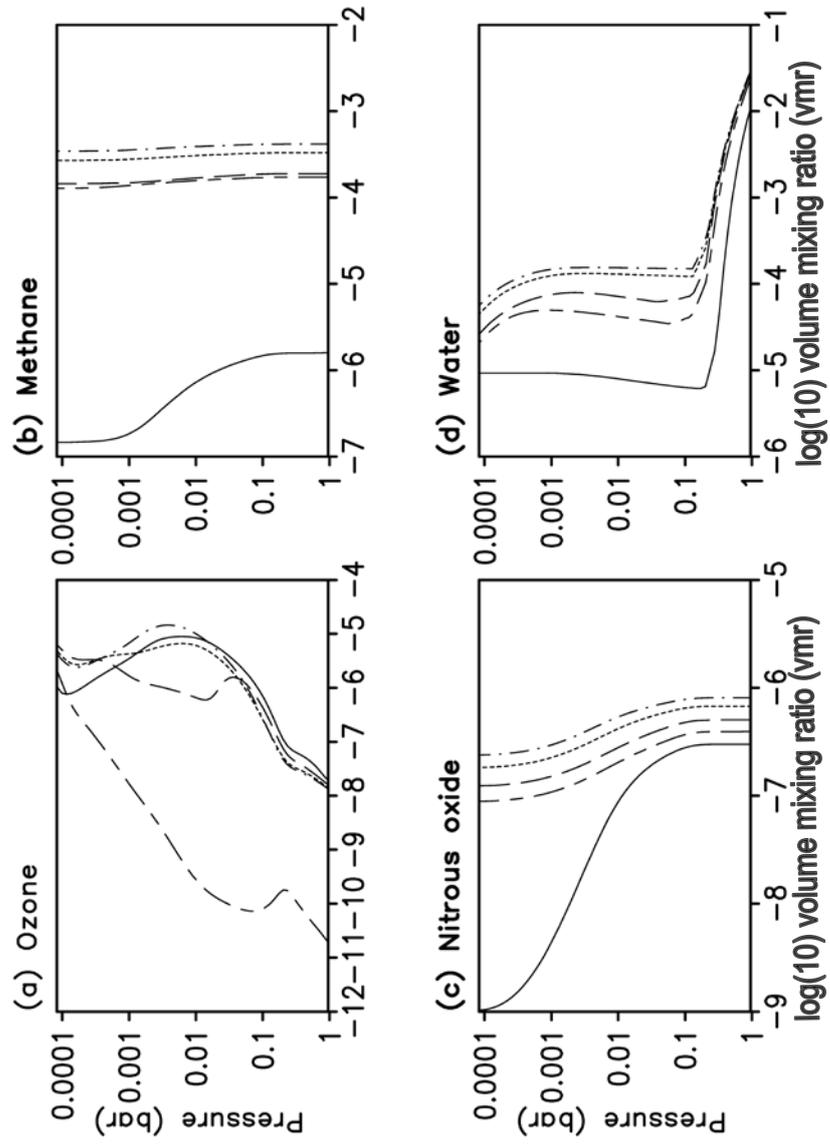

Figure 5: Biomarkers and Greenhouse Gases (LOG10 vmr)

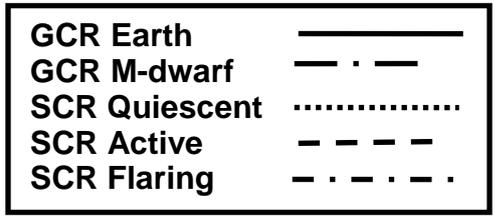

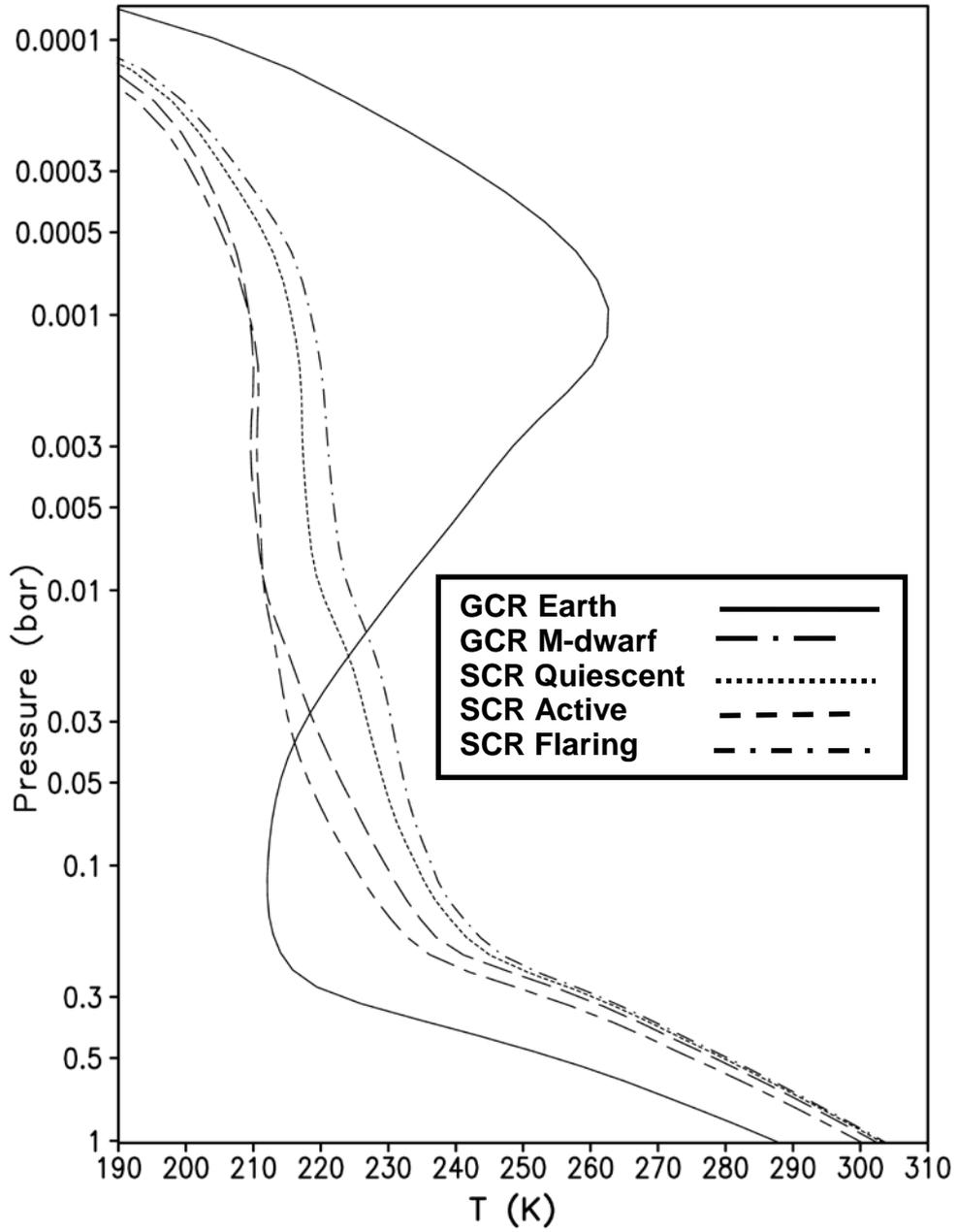

Figure 6: Atmospheric Temperature Structure

Figure 7: Thermal Emission Spectra

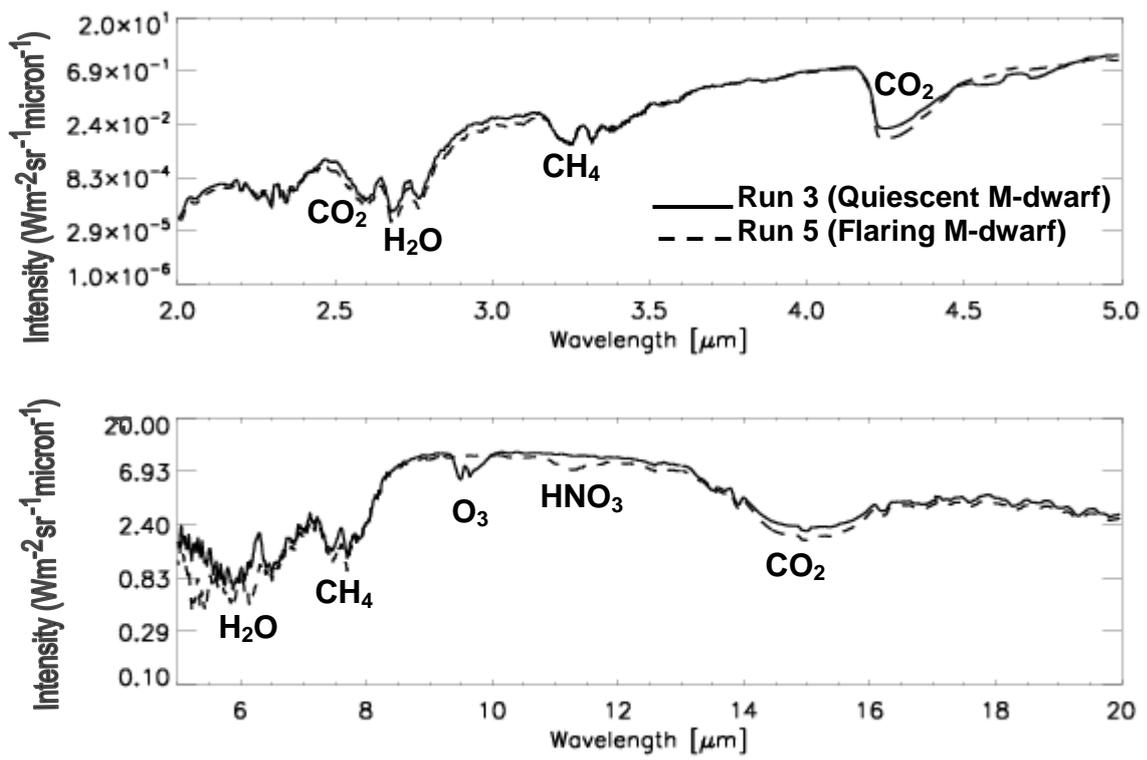